\documentstyle[12pt,epsf]{article}
\newcommand {\be}{\begin{equation}}
\newcommand {\ee}{\end{equation}}
\newcommand {\vk}{{\bf k}}
\newcommand {\vv}{{\bf v}}
\newcommand {\vn}{{\bf n}}
\newcommand {\vr}{{\bf r}}
\newcommand{\Li }{{\mbox {Li}}}
\newcommand{\arctg}{{\mbox {arctg}}}

\begin{document}

\title{Final state interaction in the production 
of heavy unstable particles}

\author{K.~Melnikov~{\thanks{ e-mail: 
melnikov@dipmza.physik.uni-mainz.de}}
~and O.~I.~Yakovlev~\thanks{{Permanent address: 
Budker Institute of Nuclear Physics, Novosibirsk 630090,
 Russia};~~ e-mail: yakovlev@dipmza.physik.uni-mainz.de;
}
\\{\em Institut fur Physik (THEP), Johannes Gutenberg
Universit\"at,}\\{\em Staudinger weg 7, D55099, Mainz, Germany}}

\maketitle

\begin{abstract}
We make an attempt to discuss in detail the effects originating from 
the final state interaction in the processes involving 
production  of unstable elementary 
particles  and their subsequent decay. 
Two complementary scenarios are 
considered: the single resonance production and the production of two 
resonances. 
We argue that part of the corrections due to the final state interaction
can be connected with 
the Coulomb phases of 
the involved charge particles; the presence of the unstable particle in 
the problem makes the Coulomb phase ``visible''. 
It is shown how corrections due to the final state interaction 
disappear when one proceeds 
to the total cross-sections.
We derive one-loop non-factorizable radiative corrections to the lowest 
order 
matrix element of both single and double resonance production.
We discuss how the infrared limit of the theories  
with the unstable particles is modified. In conclusion we briefly discuss 
our results in the context of
the forthcoming experiments on the $W^+W^-$ and the $t\bar t$ production
at LEP $2$ and NLC.
\end{abstract}

\newpage

\section {Introduction}
\par
Processes involving  production of unstable 
fundamental particles such as 
top quark, electroweak massive gauge bosons, Higgs 
boson
are now at the frontier of both theoretical and 
experimental high-energy physics.
Investigation of such processes can provide a valuable  
information about fundamental parameters
(mass, width) of the heavy unstable particles. 
 Therefore  the theory has 
 to give  reliable predictions to meet expected experimental precision.
 However it must be recognized that the accuracy of the 
theoretical description
 of the processes 
involving unstable particles and the planning accuracy of 
the measurements have been never previously combined. There exists a number 
of recent theoretical inventions which all are connected with 
the accurate description of the unstable particle in the vicinity of its 
pole: the S--matrix approach to the 
Z-pole \cite{Stuart}  and the extension of this 
scheme to a more general cases \cite{Apl},
 colour rearrangement phenomena \cite{Khoz}, etc.
\par
Currently, the main source of our 
experience in the field is the $Z$-boson physics.
However, as it has been mentioned in \cite{Apl} and 
will be quite clear from 
the forthcoming discussion, $Z$-pole description is distinguished due 
to the following points:
\begin{enumerate}
\item {$Z$-boson is neutral;}
\item {Main results have been obtained for the process 
$e^+e^- \rightarrow
Z \rightarrow f\bar f $, i.e. for the production of the 
Z-boson in the $s$-channel.}
\end{enumerate}
These features greatly simplify precise description of the
$Z$-boson  
production cross-section. 
\par
The basics  of the theoretical approach to   
the processes with the unstable particles 
can be described as follows: when two unstable 
particles are produced all Feynman graphs can be divided in two classes:
the first one includes the graphs without
interactions between
decay products of different unstable particles, while the second  includes 
such graphs in which two decay processes are not independent. Generally 
speaking the graphs  of the second type give a correction  of the relative 
order $ \alpha \Gamma/M_{l}$, where $\alpha $ is an appropriate
(depending on the process) coupling constant and 
 $M_{l}$ is the  
characteristic scale for the 
momentum flow inside the loop. For example the graphs with the 
$Z $--boson 
exchange between decay products of unstable particles 
provide $ M_{l} \approx M_Z $. Hence their 
contribution is negligible. On the other hand for the graphs with
photon or gluon exchange the scale $M_{l}$ is of the order of the 
width of the unstable particle  
$\Gamma $. Consequently such graphs 
have no additional suppression in comparison with the  factorizable 
ones \footnote{As for gluon exchange, this also verifies the use of 
the
perturbative QCD for the calculation of these contributions in the 
reactions with the top quark(s),
since $M_{loop} \sim \Gamma _t$ sets the scale on which 
$\alpha _s$ must be evaluated for these non-factorizable contributions.}.
It is not difficult to convince oneself that the dominant contribution 
comes from the soft photon or gluon region [4-6].
This contribution is a non-trivial 
function of the width of the unstable particle. 
The reason is that the soft massless particles probe the mass-shell
limit of the theory 
and it is the width of the particle that changes the mass-shell
behaviour of the resonance in comparison with the stable particle.
Hence in this problem    
(similar to the case of the  threshold production of the unstable particles
 \cite {FK}) one  
have to use resummed propagator of the unstable particle. 

The other connected physical problem which has 
attracted serious attention in the 
past years is the  photon or gluon radiation of 
the unstable particles \cite{Orr}. It turns out that in the specific 
kinematical
configuration the radiation is completely insensitive to the decay.
The reason for this can be found in the conservation of the charged
or colored currents. 

As is well known 
the soft physics generally 
``suffers'' from the cancellation between  real and virtual corrections. 
It is 
therefore quite
desirable to clarify how and when this 
cancellation occurs when unstable particles are involved and to what 
extent one can study real and virtual corrections independently. 
\par
Recently some  progress has been achieved in the 
understanding of these problems. 
It has been argued [4-6] 
that non-factorizable corrections  do not contribute 
to  
inclusive quantities, e.g. to the total cross-sections, 
when both virtual  
and real ones are taken into account.
This can be viewed as the extension of the Bloch-Nordsieck-Lee-
Nauenberg-Kinoshita cancellation for the processes involving unstable 
particles \cite{Fad2}.\footnote{ Note that in general this cancellation 
differs from the cancellation known from the 
$Z$--pole physics. This fact clearly follows 
from the ref. \cite{Fad1}.
Indeed, in describing $Z$--pole we deal only with the initial--final 
non--factorizable 
interaction which is much more simple. In general case there is also 
final--final state interaction which brings some new features to 
the problem.} 
However our analyses shows that it is not only usual 
real--virtual cancellation but something more involved. This fact 
becomes clear
if one studies the influence of the non-factorizable
radiative 
corrections on differential distributions. These 
distributions 
(say in the invariant mass of the decay products of a resonance) 
are a possible tool to investigate
fundamental parameters
of the unstable particle (for recent discussion of the $W$--boson
and the top 
physics see refs.\cite{Been}, \cite {Bern} and references there in).
It turns out  that differential distributions are 
affected by this non-factorizable interactions (to the best of our 
knowledge this fact has been first noted in 
the ref. \cite{Sum} for the specific case of the top threshold production).
\par
In what follows we evaluate $O(\alpha , \alpha _s)$ 
non-factorizable resonance
radiative corrections to the differential distributions in the invariant mass
of the unstable particle(s). If the integration over invariant masses is
performed, these
radiative corrections disappear \cite{Fad2}, \cite{Mel}. Calculating
radiative corrections to the differential cross-section we are able
to clarify the physical origin of this ``inclusive zero''.
\par
However, it turns out that {\it the shape } of the differential 
distributions  and {\it the position of the maximum} 
of the differential distribution in the
invariant mass of the resonance which could be
naively identified with the pole mass of the unstable particle 
\footnote{We remind that the lowest
order differential distribution is the common Breit-Wigner.} are affected
in the energy region slightly above threshold 
of two resonances. For the particular case of two $W$-bosons
this region is approximately $170-190$ GeV being almost the same as
the LEP2 energy region.

Absolutely deliberately we do not consider the real threshold region 
(i.e $ \sqrt{s}-2M\sim O(\Gamma )$). In this region  new physical 
phenomena 
appear (bound state formation, etc.) and our analyses would be more 
complicated there.
Our idea is to get the most clean
laboratory for the effects which are completely 
connected with the unstable nature of an appropriate particle. 
Threshold region represents a special case and has
to be discussed separately (for the top threshold production 
see \cite{Sum}, \cite{Bern}). 

Let us note that
through out the paper 
we use the Breit-Wigner with a  constant width
as the propagator for the unstable particle.
As we are concerned with the 
corrections of the order $O(\alpha)$  to the lowest order 
result we can safely use the lowest order propagators for the 
unstable particles since all modifications show up only in 
higher orders.  

Subsequent part of the paper is organised as follows: 
next section is devoted to the investigation of 
the single resonance production;
more involved scenario with the production of two 
resonances is discussed in the section 3  where all basic formulae are 
presented. In the section 4 we analyse our results in a more 
informal way. Conclusion of the whole work is given in the section 5. 
A number of helpful formulae are presented in the Appendix.

\section {Simple model}

For simplicity we start with the model describing 
scalar particles which interact with the ``photon'' field. 
Suppose
one of this particles (we call it $ W $) can decay to two other 
(electron and neutrino  for simplicity). Our $W$ particle is
produced by some neutral current (virtual photon)
together with the other stable particle ($B$- 
particle).  In such a model $W$ and $B$  have opposite
electric charges. 

In what follows we discuss reaction
$ \gamma ^* \rightarrow W^+ B^- \rightarrow e^+ \nu B^- $ 
taking into 
account $O(\alpha)$ non-factorizable radiative 
corrections\footnote{In some sense this case corresponds to the 
process $t \to W^+b$.}. 
\par
We consider this process in the center of mass frame of the virtual photon.
Then it carries the total energy $\sqrt{s}$ and the zero three momentum.
We are interested in the distribution over the invariant mass of the 
$W$ particle. In order to describe this 
distribution we introduce a parameter:
\be
\delta _W = \frac {p^2 _W - M^2 _W}{M_W} \label{off}
\ee
where $M_W$ is the pole mass of the $W$ and $p^2 _W$ is the invariant 
mass of the final $e\nu$ system.
The Born graph is shown in the fig.1. Above the production threshold 
of the $W$ particle the Born graph has the 
resonant propagator forcing produced $W$ to be almost on shell.
Non-factorizable virtual corrections are also shown in the fig.1.   
Let us first discuss  the  
graph with the $B~e$ interaction. 
Since we are interested in the corrections of the 
order $O(\alpha)$  
we have to get the resonance denominator from this graph.
Consequently the loop momentum must be small 
in order not to shift the  
$W$-particle propagator  far from the pole. From this 
it is clear that the only loop momentum 
region which can provide such "resonance" correction is the soft region,
where
one can use soft-photon approximation (cf. ref. \cite{Fad1}). 
In the soft photon approximation
the 
amplitude of this process reads 
(we use the Feynman gauge through out the paper):
\be
M_a = - \frac{4 \pi i \alpha  M_0}{D(p_W)} \int \frac {d^4k}{{(2 \pi )}^4}
 \frac {4p_B \cdot p_e}{(k^2+i\epsilon)(2p_Bk+i\epsilon)(2p_e k-i\epsilon)}
D(p_W-k).
\ee
Here
$M_0$ is the Born amplitude  and
\be
  D(p_W) = \frac {1}{{p_W}^2-{M_W}^2+iM_W\Gamma _W}
\ee
  is the 
propagator of the $W$ particle  with the finite width included
explicitly.

To perform the integration over $k$ we first integrate over zero
component of this four vector.
There are four poles in the complex plane of the $ k_0 $ variable. 
Let us 
integrate over the lower half plane of the $k_0$--complex plane. 
Then two poles have to be taken into account: one from the 
$B$ particle propagator
(`` the particle pole'') and the other one from the virtual 
photon propagator (`` the photon pole''). 
It can be seen that in contrast to the soft 
photon approximation in the QED with stable particles, in our case
the contribution due to the virtual photon pole  does not  
cancel corresponding soft photon emission immediately. 
However, their 
difference appears to be pure imaginary  and hence 
does not influence differential distributions (for more detailed discussion 
see section 4.2). Keeping 
this in mind we take into account the $B$-particle pole only.
    
The contribution due to the $B$ particle pole reads:
\be
M_a=-4\pi \alpha ~\frac {4p_B \cdot p_e }{4E_eE_B} \frac {M_0}{D(p_W)}~I,
\ee
$$
I=\int \frac {d^3k}{{(2\pi )}^3} \frac {1}{({(\vv _B \vk)}^2- {\vk}^2)~
(\vk \vv _B- \vk \vv _e -i\epsilon)} 
~D(p_W-k)|_{k_0=\vk \vv _B}.
$$
 Here the quantity  $\vv _i $ is the three velocity of the 
 $i$-th  particle.  
Using momentum conservation we get for the $W$ propagator:
\be
D(p_W-k)|_{k_0 = \vk \vv _B}= \frac {1}{M_W (\delta _W + i \Gamma _W -2 
\sqrt {s/M_W^2} \vk \vv _B)}.
\ee
In order 
to compute residual integral over $k$  it is  useful
 to exponentiate the 
propagators, introducing two different ``times''. The amplitude 
reads then:
$$
M_a= \frac {- 4 \pi \alpha 4p_B \cdot p_e M_0}{4E_eE_B M_W D(p_W)} 
\int d \tau dt \exp \{i \tau ( \delta_W +i \Gamma ) \} 
\int \frac {d^3k}{{(2\pi )}^3} 
\frac { \exp \{ i \vk \vr (t, \tau)\}}{{(\vv _B \vk)}^2- {\vk}^2}
$$
where
$ \vr (t, \tau ) $ stands for:
$$
\vr (t,\tau) = (\vv _e-\vv _B)t-2 \sqrt { \frac {s}{{M_W}^2} }
 \vv _B \tau.
$$
\par
The integral over $ \vk $ is recognized to be retarded  Coulomb 
potential of the particle moving with the velocity $ \vv _B $ and hence the 
result of $ \vk $ integration can be found in the text-books on classical 
electrodynamics:
\be
\phi(r)=-4\pi\int \frac {d^3k}{{(2\pi )}^3}    
\frac {\exp \{i \vk \vr \}}{{(\vv _B \vk)}^2- {\vk}^2}
= \frac {1}{ \sqrt {{(\vr \vn _B)}^2+(1-{\vv _B}^2)
 {\vr _\bot}^2} }.
\ee
Here $ \vn _B $ is the unit vector parallel to the B-particle velocity and
$\vr _\bot$ is the component of the vector $\vr$ transverse to the vector
$\vn _B$.
The final expression which can be obtained in this way is:
\be
M_a=  \alpha \frac {4p_B \cdot p_e }{4E_eE_B} \frac {M_0}{M_W D(p_W)}
\int d \tau dt \phi (r(t,\tau ))
\exp \{i \tau ( \delta _W +i \Gamma _W) \}. 
\ee

Let us note that the way we proceed is quite similar to the 
eikonal approximation for the high-energy scattering. It is well known in 
that case and can be proved in ours that the leading
contributions from the eikonal graphs 
to the amplitude can be summed up .
The result is the Coulomb 
phase \cite{Lip} of the wave function of the charged 
particle. Usually the Coulomb phase 
is not 
important due to its pure imaginary nature. We shall see that in our 
example
this is not the case and that {\it the residual contribution from the 
Coulomb phase survives in the final result}.
\par
Integrating the last equation over $ t $  and $ \tau $ we neglect the 
terms which are pure imaginary  and hence do not 
contribute to the differential cross-section at the $O(\alpha)$ order. 
We get then:
\be
M_a= - \alpha ~\frac { 1-\vv _e 
\vv _B}{\sqrt {{(\vv _B-\vv _e)}^2-{\vv _B}^2 
\vv ^2 _{e\bot}}} 
~i~\log \Big ( \frac {i M_W}{\delta _W + i\Gamma }\Big ). 
\ee
In this equation $\vv _{e\bot}$ is the component of the vector
$\vv _{e} $ transverse to the vector $\vn _{B} $.   
Note that the factor in front of the logarithm is nothing 
but the Lorentz boosted 
Coulomb factor $ \alpha/|\vv _1-\vv _2|$.
This factor has the following limits:  
when velocities are small it turns to   $ \alpha/|\vv _B-\vv _e|$ 
hence reproducing usual expansion parameter for the Coulomb problem 
\cite {Som}
while in the limit $ |\vv _e| \rightarrow 1 $ or 
$ |\vv _B| \rightarrow 1 $
this factor equals to $\alpha$ and 
hence appears to be independent from the kinematic of the process.

Let us now discuss the  photon exchange between $W$ 
and $B$ (see fig.1). On the first glance this graph 
does not seem to be non-factorizable correction we are interested in. 
However
gauge invariance arguments do not allow 
us to exclude this graph from the consideration. We study 
this graph in the soft-photon approximation 
neglecting the contribution of the photon pole (see sect.3.2).
Calculation is quite similar to the previous one and results in the 
following contribution to the amplitude:
\be
M_b=\alpha ~M_0 ~\frac {1-\vv _W \vv _B}{|\vv _W-\vv _B|}
~i~\log \Big ( \frac {i M_W}{\delta _W + i\Gamma } \Big ).
\ee
There are no other corrections of the non-factorizable origin
which influence differential 
distributions. For example the 
interaction of the $W$ with the electron is of the initial-final 
state interaction type \cite{Fad1} 
and hence has rather simple pole structure. The 
infrared contribution from this graph is completely cancelled by the 
corresponding real emission ( see also discussion in the
ref. \cite {Lip} for the stable particle case). 
Hence all radiative corrections 
which are of the 
non-factorizable nature and are not cancelled by the emission of the 
soft 
photons are given by the sum of two amplitudes 
presented above. The sum of this amplitudes
gives us the result for the non-factorizable corrections:
\be
M_{n/fact} = \alpha ~M_0 ~i~ 
\log \Big ( \frac {iM_W}{\delta _W + i\Gamma } \Big )
\Big ( \frac {1-\vv _W \vv _B}{|\vv _W-\vv _B|}- \frac { 1-\vv _e 
\vv _B}{\sqrt {{(\vv _B-\vv _e)}^2-{\vv _B}^2 
 \vv ^2 _{e\bot}}} \Big ).
\ee
Let us write corresponding contribution to the cross-section in the 
following form:
\be
d \sigma = d \sigma _{0}~K,~~~
K= -2~\eta ~   
\arctg \Big ( \frac { \delta _W }{\Gamma _W} \Big ),
\label{single}
\ee
$$ 
\eta = \alpha ~\Big (
 \frac {1-\vv _W \vv _B}{|\vv _W-\vv _B|}- \frac { 1-\vv _e 
\vv _B}{\sqrt {{(\vv _B-\vv _e)}^2-{\vv _B}^2 
 \vv ^2 _{e\bot}}} \Big ). 
$$
Here $d\sigma _0$ is the lowest order cross section.
The important point to be noted here is that 
in the relativistic limit for this equation
{\it the cancellation}  between contributions from  
$WB$ and $Be$ interaction occurs. This result recovers the  
``non--observability'' of the Coulomb phase. 
We could expect this compensation because in the ultra--relativistic 
limit the 
spectator (``B''-particle) 
does not distinguish transverse movement of the electron
and hence does not notice that the charge movement has been 
changed.
As the  result, the Coulomb phases
of the resonance and its decay products add coherently
to a pure imaginary quantity reproducing the non-observability of the 
Coulomb phase.
\par
Following above discussion we can 
exponentiate our results given in the eqs. (12)-(13) 
(cf. ref. \cite{Lip}) to get:
\be
 M_{n/fact} = M_{0} \Gamma (1+i\eta ) e^{i \phi _0 } {\Big ( \frac {M_W}
{\delta _W + i \Gamma _W}\Big )}^{i \eta }
\ee
where $i\phi _0$  is an imaginary phase which  is 
not relevant. 
It is straightforward to find  the contribution of this 
amplitude to the differential cross-section:
\be
d\sigma = d\sigma _0 ~\frac {\pi \eta }{sh(\pi \eta )}~
 \exp \{ -2 ~\eta~\arctg(\frac {\delta _W}{\Gamma _W}) \Big \}.
\ee

To see how this correction disappears
when we proceed to the total cross section [4-6] we 
integrate previous equation  over the range of the $ W $ masses (of 
course with the usual approximations in mind) and  get 
\be
\sigma = \sigma _0~(1 + O \Big (\alpha~ \frac {\Gamma _W}{M_W} \Big ) ).
\ee
On the one-loop level this cancellation is clearly visible 
since the $\arctg(\delta _W/\Gamma _W)$ is the odd function 
of the resonance off-shellness, while the usual Breit-Wigner is the even one. 
Therefore
their convolution is zero.
\par
It is clear however that this correction influences differential 
distributions in the invariant mass of the produced resonance. To get an 
idea of what one gets in the realistic situation let us imagine that we 
deal with the production of two equal mass resonances and the integration 
over invariant mass of one of them has been already 
performed. 
We treat the resonance which is ``integrated out'' as a stable particle.
The velocity of the "electron" is taken equal to unity. Then 
as it has been noted above the
factor $ \eta $  turns out to be independent from the
scattering angles of the final particles. In this case this
factor reads:  
\be
\eta = \frac {(1-\beta)^2}{2\beta} \alpha
\label{eta}
\ee
where 
  $$ \beta =\sqrt{1-\frac {{4M_W}^2}{s}}$$
   is the on-shell velocity 
 of the produced resonance (as far as we are not too close to the 
threshold we can use this on-shell value for the velocity).
We note  here that the $\eta$--factor in the eq.(\ref {eta}) 
very quickly goes to zero if the total energy increases.
For example, taking the mass of the resonance  equal to $ 80 $ GeV
  one sees that
for $\sqrt {s} =170 $ GeV we have $ \eta=0.65~\alpha $ and
for $ \sqrt {s} =200 $ GeV
this factor decreases up to $0.13~\alpha $  suppressing  this 
correction roughly to one order in magnitude. 
\par
However, this correction influences differential 
distributions in the invariant mass of the $W$-particle decay products moving 
the peak to the lower values of the resonance masses.
The result for the corrected distribution is 
shown in the fig.2 for the usual values of the width and the mass of the 
$W$--boson and for different energies of the process.
The position of the  maximum of the distribution
differs from the same quantity defined by the Breit-Wigner propagator.
The position of the new maximum is:
\be
\Big ({\sqrt {p ^2 _W}} \Big )_{max} - M_W \approx - \eta  \Gamma.    
\ee      
Though the pole position is not affected too much for the realistic values 
of the particle width and the coupling constant it is still comparable with
the planned accuracy of the $ W $ mass determination in the intermediate 
energy region $ \sqrt {s} \sim 170-190$ GeV. 
For higher energies this corrections are strongly suppressed hence 
having no importance from the experimental point of view. 

Let us make some comments now.

There exists the $S$--matrix  
approach  for the description of the gauge boson pole 
which was originally proposed for the description of the 
$s$--channel production of the $Z$--boson \cite {Stuart}. 
The basic idea of this method
is to start from the analytical properties of a given amplitude.
Our analysis 
shows that in order to apply this method to the {\it charge}
boson production one must claim that there is a {\it branching point} in the 
complex plane of the invariant mass of the resonance
but not a pole. Corresponding intercept of the 
branching point appears to be non-trivial function of the
kinematic of the process.
From this we conclude that a theoretical analysis of this situation
will be
more complicated and there is no straightforward extension of the 
S--matrix pole scheme to the processes involving charge unstable gauge 
boson(s)
production\footnote{This fact and a possible modification of the pole scheme
 are discussed in the ref. \cite {Apl}}. 

Our next remark concerns the top decay width. As it is clear from the exact 
expression for the $ \eta $ factor  
the non-factorizable corrections to the differential 
distributions
over invariant mass of the $W$--boson decay products are negligible
due to the small mass of the 
$b$-quark.  

Let us also outline how the
calculation of the 
radiative 
corrections to the single resonance production must be performed. 
We stress once 
more that the radiative correction which is presented in the eq.(13) 
is the only one 
which is 
usually referred as non-factorizable. As we have also traced the cancellation 
of the real emission against virtual photon poles  we can 
formulate 
the {\it practical recipe} for the calculation of 
radiative corrections to the differential cross-section of 
the reactions with one unstable 
particle: 
\par
1) The leading order differential distribution is given by the 
known formula:
 \be
  \frac {d\sigma(p^2 _W)}{dp_W^2}= \frac {\sigma _0(M_W^2)}{\pi} 
  \frac {M_W~d\Gamma (W \to e\nu)}{(p^2 _W - M^2 _W)^2+\Gamma^2 _W M^2_W} 
  \ee
where $\sigma _0$ is the on-shell cross section for the production of the 
particles $W$ and $B$, and $d\Gamma (W \to e\nu)$ is the on-shell 
differential partial 
width of the $W$-particle.

2) In order to compute $O(\alpha )$ corrections to this formula  one needs:
 \begin{itemize}    
  {\item  to substitute one-loop results for all quantities in the previous 
formula;}
 { \item to add our result for the ``Coulomb phase'' contribution (eq.13) to 
the above formula, since this is the only contribution of the order 
$O(\alpha )$
 which comes from the non--factorizable interaction. 
This prescription already takes into account partial cancellation 
of the non--factorizable real corrections against 
corresponding real ones.}
\end{itemize}

There is one 
 subtle point in the preceeding discussion. One can get 
an idea that we make a double counting, i.e. we include the soft region of 
the triangle ($bW$ interaction in the terms of the model) to the "narrow 
width approximation", while this region is also accounted in the 
calculation of the non-factorizable radiative 
corrections, which according to the recipe 
are added later by hands. We note in this respect  that  
the soft region for the ``on-shell'' triangle is completely cancelled by  
the soft emission. In this case we do not get any
contribution  from this region 
because the Coulomb phase for the stable particles is pure imaginary and 
hence disappears from the observables. 
Consequently there is no 
double counting and our recipe is simple and reasonable.  

\section {Production of two resonances}

Now we are in position to discuss 
similar problem for  the case when two  unstable particles are produced 
in an appropriate reaction.
We note that non-factorizable radiative corrections to the processes 
involving production of two resonances have been discussed
in the literature. Namely  total cross-sections [4-6] and
various distributions in the non-relativistic ( threshold) limit  \cite{Sum}
have been analysed.
As we have mentioned in the introduction, we do not discuss  
threshold region 
in what follows. 
 
 Below we calculate {\it double resonance} $O(\alpha)$ 
non-factorizable radiative corrections to the
  lowest order matrix element. Through out the paper we consistently
  neglect single-resonance and background contributions. 

Next important remark concerns logarithmic and polylogarithmic functions 
which appear in the result of this calculation. Generally we need to 
evaluate this functions in the complex plane.  Hence it is 
important to fix conventions for the cuts of this functions. All 
logarithmic and polylogarithmic functions which appear in our final 
formulae have the {\it usual} cuts i.e for the logarithms it goes from 0 
to $-\infty$ along the real axis and for the polylogarithms from 1 to
 $+\infty$ along the real axis.

For concreteness we consider the process  $ \gamma ^{*} 
\rightarrow  t \bar t \rightarrow b W^+ \bar b W^- $ as the basis for 
further discussion.  However, for the energy region sufficiently far from 
the threshold 
the results of our 
calculation  appear to be general  and are not restricted 
to a concrete process. 

It is worth to note from the very beginning that the results 
which one obtains for two resonances appear to be more complicated 
and are not so transparent from the physical point of view as compared to the 
case of a single resonance production.  If it 
is possible we try to appeal to the physical 
picture rather than huge formulae. The generic graphs 
for the non-factorizable radiative corrections 
are presented in the fig.3. 
\par
\subsection { Three-point function}
\par 
We start our consideration with the usual three-point function.
 As it is clear 
from the previous section the contribution of this graph is unavoidable 
due to the gauge invariance arguments. Corresponding amplitude diverges 
logarithmically in the soft-gluon approximation. To avoid this divergence 
we introduce the cut-off $ \Lambda $ and restrict the integration 
region to the values of the loop momenta  $ \vk ^2 \le \Lambda ^2$. 
 This regularization is not Lorentz-invariant, so we use the 
center of mass frame everywhere. The result of the calculation is Lorentz 
invariant anyhow.

As we work in the soft gluon approximation, we are interested in the
contribution from the region $k \sim \Gamma $. The natural requirement for the
cut-off $\Lambda $ is then
$$ \Gamma << \Lambda << \sqrt{s}, M_W. $$
\par
When the momentum of the virtual or real gluon is much large than $\Gamma$
we can neglect the width and the off-shellness in our formulae and work
with the usual expression for the radiative corrections.
The important point is that the usual 
radiative corrections (both real and
virtual) drop out from the 
observables in the soft gluon approximation. 
Hence we expect that the cut-off $\Lambda$ will not enter our final
formulae.

We mention here  that the complete result
 for the three point function with two unstable particles is known
in the closed form 
\cite{Bardin}. One can use this result without any 
approximations avoiding the questions associated with 
the ultraviolet divergence of the soft gluon approximation. 
However, the soft part of this triangle can 
be correctly
reproduced by  the soft gluon approximation.
  This part is necessary for our analyses providing the complementary
  gauge invariant part for the four and five point functions.

{\bf 3.1.1 Particle poles}---
The amplitude for the process under discussion corrected due to
the three--point function  in the soft-gluon approximation is:
\be
M = M_0 4 \pi \alpha _s C_F4p_1p_2 i \int \frac {d^4k}{{(2\pi)}^4}
\frac {1}{(D _1-2p_1k)~(D_2 + 2p_2k)~k^2}. 
\label{tria1}
\ee
Here  
\be
D_i={p_i}^2-{m_t}^2+im_t\Gamma _t,~~~ i=1,2
\ee
is the off-shellness parameter which we use further.
\par
$C_F$ stands for the usual colour factor:
$$ C_F=\frac {{N_c}^2-1}{2N_c}. $$
\par
Again, there are two ``particle'' and two ``gluon'' poles 
in the eq.(\ref {tria1}). We perform 
the integration over the lower half of the complex plane. 
Let us discuss the particle pole first.
\par
Taking the particle residue in the $k_0$ complex plane, introducing 
the cut--off and splitting momentum integration into parallel and 
perpendicular components with respect to the resonance velocity ( in the 
 center of mass frame two produced resonances move 
in the opposite directions) we get
$$
M_{part} = M_0\frac {\alpha _s C_F}{\pi} (1+\beta ^2)~I,   
$$
\be
I=\int \frac {dk_z}{(2 \pi)} \int 
\frac {d^2k_\bot}{{(2 \pi)}^2} \frac {1}{(D_2/E + \beta ~k_z)^2  
-{k_z}^2-k_\bot^2}~\frac {1}{(D_1 +D_2)/E+\beta ~ k_z}.
\ee
Here $E= \sqrt {s}$ is the 
total energy of the process and $\beta $ is the  
velocity of the particle with the energy $E/2$ and the mass $m_t$.

The integration over $ k_z $ is restricted to the region 
$ - \Lambda \le k_z \le \Lambda $ and the integration over $ k_\bot$ to the 
region $ k_\bot ^2 \le \Lambda ^2 - k_z ^2 $.
\par
It is seen from the eq.(22) that the integration over $ k_\bot$ is 
logarithmic and hence can be performed  immediately. Integrating 
by parts over $ k_z$ we get:
\be
M_{part}=-M_0\frac {\alpha _s C_F}{\pi} \frac {(1+\beta ^2)}{4\beta}   
\int dk_z \log\Big (\frac {D_1+D_2}{Em_t}+\frac {2\beta k_z}{m_t} \Big )
P(k_z), \label {25}
\ee
$$
P(k_z)= 
\frac {2 \beta (D_2/E+\beta k_z)}
{(D_2/E+\beta k_z)^2 - \Lambda ^2} 
-\frac {2 \beta (D_2/E+\beta ~k_z)-2k_z}
{(D_2/E+\beta k_z)^2 - k_z^2}.
$$

The simplest way to evaluate this integral  is to use 
analytical properties of the integrand in the complex plane of  
$k_z$ 
variable. The logarithmic function has branching point below 
the integration path and the singularities of the function $P(k_z)$ 
are  simple poles. Using Cauchy's theorem
we rewrite eq.(\ref {25}) as an integral over the 
half-circle of the radius $ \Lambda $ in the upper complex half-plane,
taking into account the residues where necessary. The result of the 
integration is then\footnote{We again neglect all pure imaginary quantities 
in this result.}:
\be
M_{part}=-M_0\frac {\alpha_s C_F}{\pi} \frac {(1+\beta ^2)}{4\beta}   
\Big ( 2\pi~i~\log(\xi(-1))+\pi ^2 \Big ) \label{tpart} 
\ee
where the function $\xi (x)$ will be used further through out the paper.
This function reads explicitly:
\be
\xi (x)=(1+\beta x)~\frac {D_1}{m^2_t} + (1-\beta x)~\frac {D_2}{m^2_t}.
\ee

This result shows a peculiar property --  
after the choice of the particular integration contour
we have lost the 
symmetry between two resonances, in spite of the fact that the original 
integral has such a symmetry. As we have learned before, the particle 
pole gives  the Coulomb phase, hence the absence of the symmetry in the 
particle pole contribution means 
that a part of the Coulomb phase is hidden in 
the contribution due to gluon pole. 

Let us transform the integral to the coordinate space to study 
the space-time picture. Doing so we recognize that the eikonal approximation   
provides simple deterministic picture: a 
particle is moving with the constant velocity in the constant direction, 
the ``density matrix'' is the moving delta-function. In the case when 
the particle is unstable everything is the same, except 
the normalization of the density matrix -- it is not a constant anymore. 

Starting from the expression for the amplitude presented in the 
eq.(\ref{tria1}) we 
introduce Schwinger-Fock proper time for each of the resonances. The 
integration over loop momentum reduces to the evaluation of the 
Fourier transform of the gluon propagator to the coordinate space.   
The result is well-known:
\be
\int \frac {d^4k}{(2\pi)^4} 
\frac {e^{ikx}}{k^2+i\epsilon}  
= \frac {i}{4\pi ^2}\frac {1}{x^2-i\epsilon}.
\ee
In our case the four-coordinate of the gluon propagator is the difference 
in Lorentz coordinates of the resonances:
$$
x^{\mu}= {p_1}^{\mu}\tau _1 -{p_2}^{\mu}\tau _2.
$$
In the center of mass frame 
we rewrite the result in the following way:
\be
M = -M_0\frac {\alpha _s C_F}{\pi} (1+\beta ^2)   
\int d\tau _1 d\tau _2 
\frac {\exp i(D_1 \tau _1 +D_2\tau _2)}
{{(\tau _1 -\tau _2)}^2-\beta ^2{(\tau _1+\tau _2)}^2 -
i\epsilon }.
\ee
This expression exhibits poles on the integration path. The position of 
this poles corresponds to the movement of one  particle in the field 
produced by the 
other when retardation effects are taken into account. The residues in 
these poles provide us with ( we again drop all pure imaginary quantities):
\be
M _{pole}=M_0~\frac {\alpha _s C_F}{\pi} \frac {(1+\beta ^2)}{4\beta}   
\Big ( i\pi \log ( \xi (-1) )+
i\pi \log (\xi (1))+\pi ^2 \Big ).
\ee
This expression has all  desirable symmetry properties and  
corresponds to the Coulomb phases of two resonances which they acquire 
in the field of their partners.

As the space-time picture shows that our understanding of the Coulomb
effects is still valid we proceed further and extract
the residual Coulomb phase contribution from the gluon pole.
We will not use the proper time representation systematically and continue
evaluation of the three point function in the momentum 
space.    

{\it \bf 3.1.2 Gluon poles}---
So far we have studied the ``particle'' pole contribution to the 
amplitude. Now we are in position to discuss the contribution of the gluon 
pole. The separation of particle and gluon poles in our calculation 
is useful due to the fact that the contribution from the gluon pole 
of the virtual graph is in very
close analogy to the corresponding real emission. 
Hence if we get the contribution of the virtual gluon pole
it is a matter of machinery 
substitutions to obtain the amplitude for the soft real emission. 

Taking the residue of the gluon propagator, which is located in 
the lower 
half-plane of the $k_0$ variable, we find that the integration 
over transverse component of the loop momentum is again logarithmic 
and hence straightforward. We perform one integration by parts
and 
arrive finally to the following representation for the gluon pole 
contribution:
\be
M _{g} = M_0 \frac {\alpha _s C_F}{\pi } \frac {(1+\beta ^2)}{4\beta}
( A_1 +A_2+A_3+A_4) 
\ee
where
$$
A_1=\int \limits _{-\Lambda}^{\Lambda}
dk_z \log \Big ( \frac {D_1+D_2}{Em_t}+\frac {2\beta k_z}{m_t} \Big )
\Big (\frac {\beta}
{D_1/E-\Lambda +\beta k_z}- \frac {\beta}{ 
D_2/E+\Lambda +\beta k_z}\Big ), 
$$
$$
A_2=\int \limits _{-\Lambda}^{\Lambda}
dk_z \log \Big (\frac {D_1+D_2}{Em_t}+\frac {2\beta k_z}{m_t} \Big ) 
\Big ( \frac {1-\beta}{ 
D_1/E-(1-\beta)k_z}- \frac {1+\beta}{ 
D_1/E+(1+\beta)k_z} \Big ),  
$$
$$
A_3=\int \limits _{-\Lambda}^{\Lambda}
dk_z \theta (k_z) 
\log \Big ( \frac {D_1+D_2}{Em_t}+\frac {2\beta k_z }{m_t}\Big ) \Big ( \frac {1+\beta}{
D_1/E+(1+\beta)k_z}+\frac {1+\beta}{ 
D_2/E+(1+\beta)k_z} \Big ), 
$$
$$
A_4=\int \limits _{-\Lambda}^{\Lambda}
dk_z \theta (-k_z) 
\log \Big (\frac {D_1+D_2}{Em_t}+\frac {2\beta k_z}{m_t} \Big ) \Big ( \frac {\beta-1}{ 
D_1/E-(1-\beta)k_z}+\frac {\beta-1}{ 
D_2/E-(1-\beta)k_z} \Big ). 
$$ 
Let us discuss the advantages of this representation:
evaluation of the integral $A_1$  
can be immediately reduced  
to the integration over the semi-circle of the radius $\Lambda$, 
as a consequence it will not depend on the off-shellness and the width of
the 
resonances. Therefore it will be completely cancelled by the  
real emission.
The $ A_2$ term is the extracted contribution of the particle 
pole ( hidden Coulomb phase, as it has been 
called above), the last two terms are 
specific for the gluon pole. The calculation of this integrals is 
straightforward due to the fact that  
all of them are of a polylogarithmic type. 
It is clear that we need to evaluate polylogarithms and logarithms as the 
functions of the complex argument.
We note in this respect that
all important 
points for performing logarithmic and polylogarithmic 
integrals in the complex 
plane have been discussed long ago in the ref. \cite {Hooft}.
\par
As has already been mentioned, the contribution from the $A_1$ term 
is completely
canceled by the real emission  hence we do not present it here.
The result for the $A_2$ reads:
\be
A_2= -\pi ^2-2\pi i~\log(\xi (1)).
\ee

The most involved is the evaluation of both  $A_3 $ and $ A_4 $ 
contributions. The result which one obtains after direct integration is:
\begin{eqnarray}
A_3+A_4&=&\log(\zeta ) \log\Big (\frac {(1-\beta ^2)E^2}
{|z_1||z_2|m_t^2}\Big )-\frac {1}{2}\log^2(|d_{12}|)-\log(|d_{12}|)
\log\Big (\frac {|z_1|}{|z_2|}\Big )
\nonumber \\
&+&2 \Li _2(\zeta )-\Li _2(-\zeta d_{12})-
\Li _2\Big (- \frac {\zeta }{d_{12}}\Big )
+ \frac {{\pi }^2}{2}\nonumber \\ 
&+&\frac {1}{2} {(\phi _1- \phi _2)}^2- 
(\nu _2 -\nu _1)(\phi _2 - \phi _1)-\pi (\nu _1+\nu _2).
\end{eqnarray} 
Here the following notations are used:
\begin{eqnarray}
z_1= \xi (1),~~~~ z_2= \xi(-1), \nonumber \\  
\nu _i = \mbox {arg} (z _i),~~~~ \phi _i = \mbox {arg} (D_i), \nonumber \\
d_{12} = \frac {D_1}{D_2},~~~~\zeta = \frac {1-\beta}{1+\beta}.
\nonumber
\end{eqnarray}
In the presentation of this result we split the answer into the modulus and 
the phase parts, and write each of them in a way which allows 
straightforward investigation of the $ \beta \rightarrow 1 $ limit.
\par
The result for the three-point function is then:
\be
M_{t\bar t}=M_{part}+A_2+A_3+A_4.  \label{tria}
\ee 
Here $M_{part}$ is defined in the eq.(\ref{tpart}).
\par
Now let us discuss how corresponding real emission can be obtained from 
these quantities. In particular we 
mean the interference of the gluons emitted 
by different resonances. It is straightforward to write  the 
contribution of this interference term to the differential cross-section  
in the soft-gluon approximation. Direct examination of the momentum integral 
shows that it is sufficient to perform the following modifications
in the 
result obtained for the virtual gluon pole to get a contribution due
to the real emission:
\begin{itemize}
\item {$D_1 \rightarrow {-D_1}^*$;}
\item {change the sign of the result.}
\end{itemize}
 It is important to note here that  
this transformation does not influence analytical properties 
of the amplitude, hence we can perform it in the final result. 
It is seen from the eqs.(30-31) that the virtual pole 
contribution is not invariant under this transformation, hence the real 
emission will not cancel the contribution of the virtual 
gluon pole for the three-point function. 

To demonstrate this point we study the limit 
$ \beta \rightarrow 1 $. 
It is straightforward to obtain the following from 
the eqs. (24), (30--32):
\begin{itemize}
\item {Particle pole: $ 2 \pi \phi _2 $;} 
\item {Gluon pole : $ 2 \pi \phi _1 - \pi (\phi _1+\phi _2)- 
\frac {1}{2}{(\phi _1-\phi _2)}^2+const$;}
\item { Virtual correction $=$ Particle pole + Gluon pole.}
\end{itemize}
Transition to the real emission discussed above reduces to
the transformation 
$ \phi _1 \rightarrow \pi -\phi_1$. As a result the sum 
of the real emission
and the virtual correction in the limit 
$ \beta \rightarrow 1 $ equals to:
\be
\pi (\phi _1 + \phi _2) + 2\phi _1\phi _2 + const.
\ee
The  constant term is independent from widths and off-shellnesses of the 
resonances  
and we do not present it here. The first term is the Coulomb phase of 
two resonances in the limit $\beta \to 1$  
and the second is a correlation between the phases of two 
resonances. 

\subsection {Four point function}

As a next step we consider the graphs  with 
the gluon exchange between $ t \bar b $ or $ \bar t b$. Evidently there is 
a symmetry between these two and having the result for one of them it is 
straightforward to reconstruct it for the other. For  
concreteness we study the interaction between $t$ and $\bar b$.

{\it \bf 3.2.1 Particle poles}---
We start with the discussion of the particle poles. In this 
case it is not so easy to apply direct integration discussed in respect 
with the evaluation of the three point function and we use the following trick to reduce the 
necessary amount of work: in the soft approximation the product of two 
propagators of the unstable particles 
  (cf. eq.(3)) can be decomposed as:
\be
D(p_1-k)~D(p_2+k)=\frac {1}{(D_1+D_2)+4 {\bf p} \vk} \Big ( 
\frac {1}{D_1-2p_1k} + \frac {1}{D_2+2p_2k} \Big ).
\label{dec}
\ee
Here $ {\bf p} $ is the three momentum of the produced unstable particles 
( $ {\bf p} = {\bf p} _1 = -{\bf  p} _2$).
Examining the  poles in the complex plane one finds 
that by appropriate choice of the integration contour the second term in 
this decomposition gives no particle pole  contribution 
while for the first one
it is sufficient to take the pole corresponding to the  $ \bar b $ 
propagator. In fact this decomposition leads to a mixture of the poles of 
the original expression. Hence strictly speaking the particle poles which 
are discussed below are some combinations of the original 
particle and gluon poles. 
However as we have seen during
discussing the triangle graph  particle and gluon poles are 
hardly separated when we deal with the amplitude involving two unstable 
particles. So the ``names'' here are just a matter of taste.   

Taking the residue of the $ \bar b $ propagator  we get:
$$
M_{part} =\overline {M} _0 \frac {4\pi\alpha _s C_F}{D_1} 
\frac {1-\vv _1\vv _4}{E}~I,
$$
$$
I=
\int \frac {d^3k}
{(2\pi)^3} \frac {1}{((\vv _4\vk )^2- \vk ^2) 
((D_1+D_2)/E+2 \vv _1 \vk)
(D_1/E-(\vv _4- \vv _1)\vk )}.
$$
Here $\overline {M} _0$ is the Born amplitude with the extracted 
unstable propagators,
$ \vv _1$ and $\vv _4$ stand for the top and the $\bar b$-quark 
three velocities 
respectively.
As it is clear  from this equation, the integration can be performed 
in a way similar to the case of the single resonance production. 
We introduce a proper time for each of the resonance propagators 
and  
exponentiate them. The integration over $\vk $ is then the same 
as in
the single resonance case (section 2). Finally we get:
\be
M_{part}=\overline {M} _0 \frac {\alpha _s C_F}{D_1}  
\frac {1-\vv _1\vv _4}{E}
\int \frac { d\tau d\tau _1}{\sqrt {{r_1}^2+{r_2}^2}}
\exp \Big \{i \Big ( \frac {D_1+D_2}{E} \tau +\frac {D_1}{E} \tau _1
\Big ) \Big \}. 
\ee
In what follows we denote the angle between velocity $\vv _i$ of the
particle labelled $i$
and the top quark velocity $\vv _1$ 
by $\theta _i$.
Then $r_1$ and $r_2$ are:
$$
r_1=2\beta \cos \theta _4 \tau - (1-\beta \cos \theta _4)\tau _1
$$
and
$$
r^2_2= \frac {{m_4}^2}{{E_4}^2} \beta ^2 \sin ^2\theta _4{(2\tau +\tau_1)}^2. 
$$
Next we make the change of the variables: $ \tau =
\lambda x$ and $\tau _1 =\lambda (1-x)$ with $ 0\le x\le 1$ and
$0\le \lambda \le \infty$. Integrating further over 
$\lambda$ we obtain:
$$
M_{part}=\overline {M}_0 \frac {i\alpha _s C_F}{D_1} (1-\vv _1\vv _4) 
\int \limits _{0}^{1}   \frac {dx}{D_1+D_2x}
\frac {1}{\sqrt {\Big [ (1-\beta x_4)-x(1+\beta x_4) \Big ]^2+\mu ^2{(1+x)}^2}}.
$$
 We denote $ \cos \theta _4 $ as $x_4$ and :
$$
\mu ^2= \frac {{m_4}^2}{{E_4}^2}\beta ^2(1-x_4^2).
$$
Examining  previous equations we see that the leading 
term 
under the square route can go through zero within the integration region 
if $ x_4 \ge 0 $.In this case this ``would be'' divergence is regularized by 
keeping the mass of the light particle finite. This means that the 
divergence is collinear. Actually this divergence can appear only if the 
"mass" of the gluon is zero, i.e. when the gluon pole in the original 
expression is taken. Hence the appearance of this divergence in the particle 
pole means that decomposition  of the unstable propagators (see eq.(34))
which we 
use for the evaluation of this graph has really mixed gluon and particle 
poles of the original expression in a nontrivial way.
\par
We insert the identity $1=\theta (x_4)+\theta (-x_4)$ 
inside the integral.
After this we get:
\be
M_{part}=\overline {M}_0 \frac {i\alpha _s C_F(1-\vv _1\vv _4)}{D_1\xi (x_4)m_t^2}  
\Big ( \theta (-x_4)~A+\theta (x_4) ~B \Big ), 
\ee
$$
A=\log (\frac {D_1+D_2}{D_1})+ \log(1-\beta x_4)-\log(-2\beta x_4),
$$
$$
B=\log\Big (\frac {\xi (x_4)m_t^2}{D_1+D_2}\Big )-
\log \Big (\frac {\xi (x_4)m_t^2}{D_1} \Big )
+\log(1-\beta x_4)+\log(2\beta x_4)
$$
$$
-\log\big [\beta ^2(1-{x_4}^2) \big ] +\log\Big (\frac {{E_4}^2}{{m_4}^2}\Big ). 
$$
\par
As it has been mentioned before 
and is quite clear from the above equation there 
are collinear logarithms associated with the massless $b$-quark in the 
final state. We discuss below 
(see section 3) how collinear logarithms cancel in the final result.  

{\it \bf 2.2.2 Gluon pole}---
Using decomposition eq.(34) for the resonance  propagators
and performing the integration over the contours discussed above 
we are forced to take the lower pole of the
gluon propagator  for the first term in the 
decomposition and the upper one for the second. Performing the 
integration over the modulus of the three-momentum we 
obtain the 
following representation for the amplitude:
\be
  M_g= \overline {M}_0 \frac {\alpha _s C_F}{\pi D_1m_t^2} 
 \frac {1-\vv _1\vv _4}{2} 
(J_1+J_2).
\ee
Where $J_1$ and $J_2$ are:
\be
J_1=\int \frac {d^3 \vn _k}{2\pi \xi (x) 
(1-\vn _k\vn _3)}
\Big (\log\Big (\frac {2\beta x+i\epsilon}{1-\beta x} \Big )
+\log\Big (\frac {D_1}{D_1+D_2} \Big ) -i\pi \Big ),
\ee
\be
J_2=-\int \frac {d^3 \vn _k}{2\pi \xi (-x) (1-\vn _k\vn _3)}
\Big (\log\Big (\frac {2\beta x+i\epsilon}{1-\beta x} \Big )
+\log\Big (\frac {D_2}{D_1+D_2} \Big ) -i\pi \Big ).
\ee
Here $\vn _k$ is the unit vector, by $x$ we denote 
 $\cos \theta _k =\vn _k\vn _1$ and $\vn _i$ is the unit
vector parallel to the velocity of the particle $i$. 

The integration over azimuthal angle  
is easily performed using the following equation:
\be
\int \limits_{0}^{2\pi} \frac {d\varphi}{1-\vn _k\vn _i}=
\frac {2 \pi }{\sqrt {{(\cos \theta _k-\cos \theta _i)}^2}}.
\ee
This equation exhibits collinear singularities which appear when the 
momentum of the gluon is parallel to the momentum of the (anti)quark. 
We regularize them keeping the mass of the light particle 
    in the 
  singular terms. The exact formula reads:
\be
|x-x_i| \to \sqrt {{(x-x_i)}^2+\frac {{m_i}^2}{{E_i}^2}
(1-{x_i}^2)}.
\ee
Finally  changing the sign of the integration variable in the eq.(39)  
$\cos \theta _k \to -\cos \theta _k$ we get:
\be
J_1+J_2=\int  \limits _{-1}^{1} \frac {dx}{\xi (x)~\sqrt {(x-x_4)^2}}   
\Big ( \log\Big (\frac {1+\beta x}{1-\beta x}\Big )
+\log\Big (\frac {D_1}{D_2}\Big ) +i\pi \theta (-x)
-i\pi \theta (x) \Big ). 
\ee

It is rather straightforward to calculate the integral in the last 
equation. We split the integration region into two parts 
to rewrite the 
square root correctly and use a partial fractioning to obtain 
Spence--like integrals. 

It is quite useful here to  
examine a part of the 
previous expression which contains $\theta $-functions. 
The evaluation is straightforward. The result is:
\be
M_{g \theta}= \overline  {M} _0 \frac {\alpha _s C_F}{\pi} 
  \frac {(1-\vv _1\vv _4)}{2D_1\xi (x_4)m_t^2}~A,
\ee
$$
  A=(i\pi \theta (-x_4)-i\pi \theta (x_4))~A_1
+2i\pi \theta (x_4)~A_2+2i\pi \theta (-x_4)~A_3. 
$$
Where $M_{g\theta}$ is the piece of the gluon pole part proportional to 
the $ \theta$-functions and 
$$
A_1=\log\Big (\frac {\xi (x_4)}{\xi (-1)} \Big)
+\log\Big (\frac {\xi (x_4)}{\xi (1)} \Big )+
\log\Big (\frac {4{E_4}^2}{{m_4}^2} \Big ),
$$
$$
A_2=\log\Big (\frac {D_1+D_2}{\xi(-1)m_t^2}\Big )+
\log\Big ( \frac {1+x_4}{x_4} \Big ),
$$
$$
A_3=-\log\Big (\frac {D_1+D_2}{\xi(1)}\Big )-
\log\Big ( \frac {1-x_4}{-x_4} \Big ).
$$

If we sum the particle pole contribution elaborated above and the 
``$\theta $ ''-part of the contribution due to the 
gluon pole  the result appears to be 
simple and all $\theta$-functions drop out.
\be
M_{part}+M_{g\theta}=\overline {M} _0 \frac {\alpha _s C_F}{\pi} 
\frac {(1-\beta x_4)}{2}
\frac {i\pi}{D_1\xi (x_4)}~K,
\ee
$$
K= 2 \log \Big (\frac {\xi (x_4)}{D_1}\Big ) +
\log\Big (\frac {\xi (1)}{\xi (-1)}\Big )
+2\log\Big (\frac {1-\beta x_4}{\beta(1-x_4)}\Big )+L_4.
$$
In this equation $L_4$ stands for 
       $$L_4=\log \Big (\frac {{E_4}^2}{{m_4}^2}\Big ).$$
This notation will be used further.
\par
To evaluate  remaining 
contributions due to gluon pole it is convenient to use  
additional functions introduced in the Appendix. Finally
we get the following result for the 
four-point function:
\be
M_{t\bar b}=\overline {M} _0 \frac {\alpha _s C_F}{2\pi} 
\frac {1-\beta x_4}{D_1\xi (x_4) m_t^2}~A_{t\bar b},
\ee
$$
A_{ t\bar b} = A_1+A_2 +A_3,
$$
$$
A_1 = -F_1(-D_0,\beta|x_4)+
F_1(-D_0,-\beta|x_4)  
+F_1(x_4,\beta|x_4)-F_1(x_4,-\beta |x_4),
$$
$$
A_2 = 
\log\Big (\frac {D_1}{D_2}\Big ) \Big [ F_2(x_4|x_4)-F_2(-D_0|x_4) \Big ], 
$$
$$
A_3= i\pi \Big \{ 2 \log \Big ( \frac {\xi (x_4)m_t^2}{D_1}\Big ) +
\log\Big (\frac {\xi (1)}{\xi (-1)}\Big )
+2\log\Big (\frac {1-\beta x_4}{\beta(1-x_4)}\Big )+L_4 \Big \}.
$$
In this equation we denote:
$$
D_0=\frac {D_1+D_2}{\beta(D_1-D_2)}.
$$

It is also straightforward to consider gluon pole of the original
matrix element (without decomposing resonance propagators
eq.(34)). 
We need it due to the study of the bremsstrahlung
integral, namely the interference of the gluon radiation
from $t$ and $\bar b$ quarks. The calculation is similar to the one 
described above. Finally we get:
\be
M_{g}=\overline {M} _0 \frac {\alpha _s C_F}{2\pi} 
\frac {1-\beta x_4}{D_1\xi (x_4)m_t^2}~A_g, 
\ee
$$
A_g = A_1 +A_2,
$$
$$
A_1=
-F_1(-D_0,\beta|x_4)+
F_1(-D_0,-\beta|x_4) 
+F_1(x_4,\beta|x_4)-F_1(x_4,-\beta |x_4),
$$
$$
A_2=
\Big ( \log\Big ( \frac {D_1}{D_2} \Big ) + i\pi\Big )
\Big (F_2(x_4|x_4)-F_2(-D_0|x_4) \Big ).  
$$
Corresponding bremsstrahlung integral can be obtained from the previous 
equation by the standard change ( cf. discussion after eq.(32)) 
in the relevant piece of the differential cross-section.

\subsection{ Five-point function}

We are finally left with the last non-factorizable graph which 
corresponds to  the interaction between $b$ and $\bar b$. 
We again calculate a contribution due to  
particle and gluon poles separately.

{\it \bf 3.3.1. Particle poles}---
Evaluation of the particle pole proceeds in a way similar to the one
which has been  used 
for the four-point function. We use decomposition 
for the propagators (eq.(34)) and then choose appropriate 
contour for the integration. The integral naturally splits into two 
pieces which represent the movement of a system 
of particles in 
the Coulomb field produced by a quark or
an antiquark. These two pieces are symmetric and complementary to each other.
\par
Let us examine one of them.
We exponentiate the propagators to obtain the Coulomb-like three 
momentum integral. As we have one more propagator here in comparison with 
the four-point function we need  to introduce three ``times'' instead of 
two: 
\be
M_{part,4}=\overline {M}_0~4\pi i\alpha _s C_F\frac {(1-\vv _3 \vv _4)}{E^2}
~I,
\ee
$$
I =  
\int d\tau d\tau _1dt \exp \Big \{i\Big (\frac {D_1+D_2}{E}\tau 
+ \frac {D_1}{E}\tau _1\Big ) \Big \}
\int \frac {d^3k}{{(2\pi )}^3} 
\frac {\exp \Big \{ i\vk \vr \Big \} }{(\vv _4\vk)^2- \vk ^2 }
$$
where $\vr $ stands for the following vector:
\be
\vr =2\vv \tau+(\vv -\vv _4)\tau _1+(\vv _3-\vv _4)t.
\ee
Here the quantity $\vv $ is the on-shell velocity of the top quark.
Integrating this equation over $\vk$ we get:
\be
M_{part,4}= i\alpha _s C_F \frac {(1-\vv _3\vv _4)}{E^2}  
\overline {M}_0\int \frac {d\tau d\tau _1dt}{\sqrt {(\vn _4 \vr )^2+\vr _{\bot}^2}}
\exp \Big \{ i \Big (\frac {D_1+D_2}{E}\tau + \frac {D_1}{E}\tau _1 \Big )
\Big \}.
\ee
Here $\vn _4$ is the unit vector parallel to the velocity of the particle 
$4$ and $\vr _\bot $ is the component of the vector $\vr $ perpendicular to 
the vector $\vn _4$.

The second term $( M_{part,3})$ can 
be obtained from the eq.(52) by the following set of 
substitutions:
\be 
  \vv _3 \to -\vv _4, ~~~~ \vv _4 \to -\vv_3, 
~~~ D_1 \to D_2, ~~~D_2 \to D_1. 
\ee

The evident intention then is to perform the integration over $t$. 
The integral 
appears to be logarithmically divergent on the upper limit. This reflects 
the fact that infrared singularities of the five-point function can not 
be completely regularized by the virtualities and widths of the 
unstable particles.
However,  we anticipate that 
the above divergence corresponds 
to the 
Coulomb phase of the $b$ quark in the field of 
the antiquark $\bar b$.
Hence we expect that this divergence is pure imaginary and drops  
from the  observable 
quantities (as 
it occurs in the infrared limit of the ``stable'' theory \cite{Suura}). 
This expectation is
verified by 
direct calculation. Below we omit this 
infinite piece from all expressions.
\par
The integration in the eq.(49) is then straightforward. 
We do not present its results
because it is much more reasonable to present the 
sum of the particle pole and the ``$\theta$''-terms 
from the gluon pole.  

{\it \bf 3.3.2 Gluon poles}---
Let us discuss the contribution due to gluon poles. 
As we have chosen
appropriate contour to evaluate particle 
poles we are forced to take the lower and the upper poles in the 
gluon propagator for the first 
and the second term in the eq.(\ref{dec}) 
respectively. As in the case of the four-point function 
we perform the integration over the modulus of the three 
momentum and get:
\be
M_g=-\overline {M}_0 \frac {\alpha _s C_F}{2\pi}(1-\vn _3\vn _4)
\int 
\frac {d^2\vn _k}{(1-\vn _k\vn _3)(1-\vn _k\vn _4)}
\Psi (D_1,D_2,\cos \theta _k). 
\ee

The function $ \Psi $ can be written in the following way:
\be 
\Psi(D_1,D_2,x)=\Psi _0(D_1,D_2,x)+\Psi _1(D_1,D_2,\beta,x)+
\Psi _1(D_2,D_1,-\beta,x)+\Psi _{\theta}
\ee
where:
$$
\Psi _0(D_1,D_2,x)=\frac {1}{D_1D_2}
\Big ( \log\frac {m_t^2}{\epsilon E}-i\pi \Big),
$$
$$
\Psi _1(D_1,D_2,\beta,x)=
\frac {1-\beta x}{D_1 \xi (x)m_t^2}\log\Big (\frac {D_1}{(1-\beta x)m_t^2}
\Big ),
$$
$$
\Psi _{\theta}= 
\frac {2 i\pi \beta x (\theta (x)-\theta (-x))}{m_t^2\xi (x)(D_1+D_2)}.
$$
The parameter $\epsilon$ in this equation is the infrared cut-off.
The first term in the previous equation does 
not contribute to the observable quantities. 
Indeed, the infrared log is canceled by the real
emission while the $ i\pi$ term is pure imaginary and hence does not 
interfere with the Born amplitude.
\par
Next we evaluate the integral in the eq.(51).  
As the  function $\Psi$ does not depend on the azimuthal angle, we  
calculate the following integral:
\be
I_{34}=\int \limits _{0}^{2\pi} 
\frac {d\varphi}{2\pi(1-\vn _k \vn _3)(1-\vn _k\vn _4)}.
\ee
Direct integration gives:
\be
I_{34}=\frac {A_{34}}{(1-\vn _3\vn _4)(1+x_{34})(x-x_a)(x-x_b)},
\ee
$$
A_{34}=\frac {N_3-xK_3}{|x-x_3|}+\frac {N_4-xK_4}{|x-x_4|}, 
$$
$$
x_{a(b)}=\frac {\cos \theta _3+\cos \theta _4 \pm i~\sin \theta _3 
\sin \theta _4 \sin \varphi _{34}}{1+\cos \theta _{34}} ,
$$
$$
N_3=1-\cos \theta _{34} -\cos \theta _3 (\cos \theta _3 -\cos \theta _4),
$$
$$
N_4=1-\cos \theta _{34}-\cos \theta _4 (\cos \theta _4 - \cos \theta _3 ),
$$
$$
K_3=\cos \theta _4 -\cos \theta _3 \cos \theta _{34},
$$
$$
K_4=\cos \theta _3 -\cos \theta _4 \cos \theta _{34}.
$$

Here we denote by $x= \cos \theta $, by $x_{i}= \cos \theta _i $ and
by $x_{34} = \cos \theta _{34} $. Here the angle
$\theta _{34} $ is the angle between the vectors $\vn _3 $ and 
$\vn _4$.

As it is seen from this equation, the result of the 
azimuthal integration is 
a  rational function of the $ \cos(\theta ) $. The remainder  
of the integrand consists of logs and constants, hence
it is quite clear 
that the integration can be performed in terms of the Spence 
functions and logarithms.
\par
The other point is that  divergence which occurs for $ x=x_{3,4}$
is the collinear one and hence its regularization 
is clear. Explicit formula reads:
$$
|x-x_i| \rightarrow \sqrt {{(x-x_i)}^2+\frac {{m_i}^2}{{E_i}^2}
(1-{x_i}^2)}.
$$

We write eq.(54)  in the following way:
\be
I_{34}=\frac {2\pi}{1-\vn _3\vn _4}(I_3(x)+I_4(x))
\ee
where
\be
I_3=\frac {N_3-xK_3}{(1+x_{34})(x-x_a)(x-x_b)|x-x_3|}.
\ee
\par
Let us study the $\theta$-terms of the $\Psi$ function 
and show how they cancel against  
corresponding parts of the particle pole. As both particle and gluon 
poles are  naturally splitted into two terms (3 and 4) we
present them 
separately.
Further evaluation is straightforward. The sum of the 
particle pole contribution and the $\theta $-terms of the gluon pole reads:
\be
M_{41}=M_{part,4}+M_{g,\theta,4}=
-\overline {M}_0 \frac {i\alpha _s C_F}{2}~K,
\ee
$$
K=
\frac {2}{D_2D}\log\Big (\frac {D}{im_t^2}\Big )
-\frac {2}{D_2D_1}\log\Big (\frac {D_1}{im_t^2}\Big )+
$$
$$
\frac {2(1-\beta x_4)}{D_1\xi (x_4)m_t^2}\log\Big (1-\beta x_4 \Big )-
\frac {2(1+\beta x_4)}{D_2\xi (x_4)m_t^2}\log\Big (\frac {D}{D_1} \Big )
+
$$
$$
\frac {1}{D} \Big \{ R_{4\xi}\Big ( 2\log\Big (\frac {\xi(x_4)}{D}\Big )
+\log\Big(\frac {\xi (1)}{\xi (-1)} \Big ) \Big )+
$$
$$
\sum _{i=\pm}R_i\Big (\log(1-x_i)-\log(-1-x_i)-2\log(-x_i)+
2\log(x_4-x_i)\Big ) +
$$
$$
R_4\Big (2~\log(\beta)+2\log(1-x_4)-L_4\Big ) \Big \}.
$$
and
\be
M_{31}=M_{part,3}+M_{g,\theta,3}=
-\overline {M}_0 \frac {i\alpha _s C_F}{2}~K,
\ee
$$
K=
\frac {2}{D_1D}\log\Big (\frac {D}{im_t^2}\Big )-
\frac {2}{D_2D_1}\log\Big (\frac {D_2}{im_t^2}\Big )+
$$
$$
\frac {2(1+\beta x_3)}{D_2\xi (x_3)m_t^2}\log\Big (1+\beta x_3\Big )-
\frac {2(1-\beta x_3)}{D_1\xi (x_3) m_t^2}\log\Big (\frac {D}{D_2}\Big )-
$$
$$
\frac {1}{D}\Big \{
R_{3\xi}\Big (2\log\Big (\frac {\xi(x_3)m_t^2}{D}\Big )
+\log\Big (\frac {\xi (-1)}{\xi (1)}\Big ) \Big)-
$$
$$
\sum _{i=\pm}R_i \Big (\log(1-x_i)-\log(-1-x_i)+2\log(-x_i)-
2\log(x_3-x_i)\Big )+
$$
$$
R_3\Big (2\log(\beta)+2\log(1+x_3)-L_3\Big ) \Big \}. 
$$
We denote $D=D_1+D_2$ in the above expression .
The exact expressions for the  quantities $R_i$ can be found in the 
Appendix. 

Then we are left with  the integration of the $\Psi _1$
function.  The result of the 
integration is:
$$
M_{42}=
-\overline {M}_0 \frac {\alpha _s C_F}{2\pi}
\Big [~\frac {(1-\beta x_4)}{D_1\xi (x_4)m_t^2}
\Big (\log\Big (\frac {4{E_4}^2}{{m_4}^2}\Big )
\log\Big (\frac {D_1}{m_t^2} \Big )-F_1(x_4,-\beta|x_4)\Big )-
$$
$$
\sum _{i=\pm} \frac {(1-\beta x_i)}{2D_1\xi (x_i) m_t^2}
\Big (\log\Big (\frac {D_1}{m_t^2}\Big )F_2(x_i|x_4)-F_1(x_i,-\beta|x_4)
\Big )-
$$
$$
\frac {R_{4\xi}}{D_1+D_2}(\log\Big (\frac {D_1}{m_t^2} \Big )F_2(-D_0|x_4)
-
F_1(-D_0,-\beta|x_4) \Big ]+ 
$$
\be
(D_1 \rightarrow D_2,\beta \rightarrow -\beta ). 
\ee
Similar term ($M_{32}$)
which corresponds to the particle 3 can be then 
obtained by 
the direct substitution $3 \rightarrow 4$ in the eq.(59).
Our final result for the radiative correction due to the $b\bar b$
interaction can be 
constructed from the above quantities:
\be
M_{b\bar b}=M_{41}+M_{42}+M_{31}+M_{32}.   \label{five}
\ee

Finally we present the
contribution of the ``true'' gluon pole
(i.e. without decomposition of the resonance propagators
eq.(34)) of the virtual five point function:
\be
M_g=M_{42}+M_{32}+M_{34}
\ee
$$
M_{34}= -\overline {M}_0\frac {i\alpha _s C_F}{2} 
\Big [~\frac {(1-\beta x_4)}{D_1\xi (x_4)m_t^2}
\log\Big (\frac {4{E_4}^2}{{m_4}^2} \Big )- 
\sum _{i=\pm} \frac {(1-\beta x_i)}{2D_1\xi (x_i) m_t^2}
F_2(x_i|x_4)
$$
$$
+\frac {R_{4\xi}}{D}F_2(-D_0|x_4) \Big ] + (4 \to 3).
$$
This concludes our evaluation of the five-point function.

\section { Analyses of the general formulae }

So far we have derived general formulae 
for the double resonance radiative corrections 
to the matrix element of the production of two resonances. 
Here we  
discuss some general properties of the obtained formulae.

\subsection {Collinear singularities}

As it is clearly seen from the above formulae each of the separate 
contributions to the non-factorizable radiative corrections exhibits  
collinear logarithms. Normally these logarithms are cancelled against 
the real emission. 
Let us note that T.~D.~Lee and 
M.~Nauenberg \cite{Lee} have used quite general approach 
to prove the 
absence of the similar divergencies in any quantum mechanical system.
The basis  for the proof is the existence of the unitary 
S-matrix.
As is well-known from the work by M.~Veltman \cite{Vel}, it 
is indeed 
possible to construct the unitary S-matrix in the field theory with the 
unstable particle. Hence the arguments of the ref. \cite {Lee} 
must apply also here. 
However 
it is necessary to clarify the level of the inclusiveness which
is necessary for this cancellation to occur when 
unstable particles are considered.

For this aim we extract  all the terms which are 
singular in the limit $m_i \to 0$, $i=3,4$ from the above formulae 
and calculate their
contribution to the total cross-section. In spite the fact that these terms 
are quite complicated in the individual graphs the sum of all these 
contributions appear to be very simple. We first write its contribution to 
the differential cross section:
\begin{equation}
\frac {d\sigma _{col}}{d\sigma _{0}} =2~\frac {\alpha _s C_F}{2\pi}~ 
\mbox { Re} \Big \{ i\pi 
\Big ( \frac {D_2}{D} L_4 + \frac {D_1}{D} L_3 \Big ) \Big \}.
\label {col}
\end{equation}
We remind that the quantities $L_3$, $L_4$ are defined 
by the eq.(42).

Let us discuss now the properties of this equation. First we note that
the source of this large logarithms are the virtual contribution
due to the five--point function.
Of course there are collinear logarimths also in the real interference but 
these are cancelled against similar pieces in the virtual corrections.

We can also reexpress the terms in the eq. (\ref {col}) to indicate
exactly the mass singularities which we find in this case:
\begin{equation}
\frac {d\sigma _{mass}}{d\sigma _0} = \frac {\alpha _s ~C_F}{2\pi} 
\mbox { Re} \Big \{ i\pi 
\Big ( \frac {D_2}{D} - \frac {D_1}{D} \Big ) \Big \} 
\log \frac {m_3^2}{ m_4 ^2}.
\label {col1}
\end{equation}

All other terms which have been dropped in the transition from the 
eq. (\ref {col}) to the eq. (\ref {col1}) are smooth in the limit when 
the masses $m_i,~~i=3,4$ go to zero.

We see therefore that if the masses of light particles in the final state 
are equal then we do not get any mass singularities. This is the case for 
instance for the reaction $e^+e^- \to t\bar t \to W^+W^- b\bar b$.
However such mass singularities will appear in the reactions like
$e^-e^+ \to W^+W^- \to e^-\bar \nu _e \mu ^+ \nu _\mu $.
They drop out if the sum of charge conjugate 
channels is considered simultaneously -- for instance 
$e^-e^+ \to W^+W^- \to e^-\bar \nu _e \mu ^+ \nu _\mu $ and
$e^-e^+ \to W^+W^- \to \mu ^- \bar \nu _\mu e^+ \nu _e$.
In any case if the integration over invariant masses of the produced 
resonances is performed, 
these singularities drop out from the observable cross section.

\subsection {Real emission and virtual gluon poles}

We discuss here how the cancellation of the  
real emission and the virtual corrections occurs when 
unstable particles are 
produced. We begin with the single resonance 
production (cf. section $1$).

First we examine  $Be$ interaction (in terms of the section 1 ).
It is straightforward to write the cross-section for the real emission 
and the contribution of the virtual photon pole to the cross-section  in 
the soft photon approximation:
$$
d {\sigma _{\gamma }}^{virt}+d {\sigma }^{emiss} =
4 \pi \alpha {|M _{Born}|}^2 4 p_2 p_3 \int \frac {d^3k}{{2 \pi }^3 2|\vk|}
2 \mbox {Re} \Big \{ \frac {D(p_1+k)+D(p_1-k)}{(2p_3k)(2p_2k)D(p_1)}
\Big \}.
$$ 
It is seen from this expression that gluon momentum enters  
the propagator of the unstable particle with different 
signs in virtual and real corrections.
This is the illustration of the statement in the ref. \cite {Fad1} 
where the authors claim 
that the 
cancellation is not local in the momentum space in contrast to the usual 
situation. However the above  expression is well-defined and we can evaluate
it explicitly. The 
result of this calculation appears to be pure imaginary and hence does not 
contribute to the cross-section. The same situation also occurs 
for the usual triangle graph with one unstable particle. 
However the case with two resonances appears to be much more unusual.

As is well known the usual thing in dealing with the soft limit of the 
Feynman graphs is the cancellation between real and virtual corrections. 
The 
essence of this cancellation is the fact that the
 particle movement is not affected by emission and absorbtion 
of soft massless quanta.
Therefore the probability of a 
process remains the same. The piece of the virtual corrections that cancels 
real 
emission is the residue of the massless gauge boson propagator ( photon 
or gluon ). 
\par
This simple remark verifies  similar cancellation in the case when the 
integration over invariant masses of the unstable particles has been 
performed. In this case, as it is clear from our consideration, we 
effectively recover 
the situation with the stable particles. However the 
differential distributions represent a different case.
\par
Explicit investigation of the contribution due to the 
gluon pole  from the 
virtual correction and the real 
emission shows  that they cancel each other in a non-trivial way. Let us 
fix the off-shellness eq.(\ref{off}) of one of the resonances 
$\delta _1$. 
Then the virtual 
gluon pole contribution calculated for the off-shellness of the other 
resonance $ \delta _2 $ cancels the real emission for the off-shellness 
$ - \delta _2 $. The reason is  that for 
negative values of $ \delta $ the particle is more likely to absorb gluons 
(the particle prefers to make its invariant mass larger)
while for positive $ \delta $'s the situation is  
opposite; exactly on the mass shell $\delta =0$ there is no difference. 
This shows that in the case of the unstable 
particle we have one more degree of freedom -- the invariant mass 
which is sensitive to the 
soft ($k \sim \Gamma $) emission and absorbtion.
Averaging over invariant masses we ``lose'' this degree of 
freedom (technically non-local cancellation in the space of the invariant 
masses occurs), but when the distribution in the invariant mass
is studied we meet some unusual properties. 
 
\section {Conclusions}

We have derived general formulae for the non-factorizable radiative 
corrections to the invariant mass distributions  
for both single and
double resonance production. We find these corrections to be important
for the accurate description of this distribution  
in the vicinity of the resonance peak. 

Our approach is motivated by the observation that non-factorizable  
corrections to  
the Born amplitude are governed by the soft
limit in order to give resonant contributions. 
This fact justifies the use of the 
soft photon (gluon) approximation for this problem.  
As usual the soft photon approximation provides  
universal results in the sense that they are not restricted to a concrete 
process.

The gauge-invariant current can only be 
constructed if one takes into 
account both the current of the resonance and the current of its decay 
products. 
Gauge invariance is responsible for the cancellation of the 
whole effect for high energies and the most probable kinematical 
configuration, i.e. when the charge decay products follow the direction of
motion of the  resonance. 

We hope that our study provides better understanding 
of the structure of the 
infrared limit of 
the theories with the unstable particles.  We note that 
the usual cancellation between soft real and 
virtual corrections is not complete
even in the well known theories like QED:
in fact the ``photon'' poles from the virtual corrections 
cancel the real emission, while the ``particle'' poles (which also give 
infrared divergencies) appear to be pure imaginary and physically correspond 
to the Coulomb phase \cite{Suura}, \cite{Lip}.

In the case when we deal with the unstable particles 
the ``particle'' poles 
provide 
non-vanishing corrections to the observable quantities. 
The origin of this correction is 
very simple: the decay of the resonance accidentally changes the movement 
of the charge and hence destroys a coherence necessary to acquire ``proper''
Coulomb phase. Dealing with the Born amplitudes describing resonance 
production we can recognize that the integration over invariant masses 
of the resonance restores the ``stable particle scenario''. 
As for the non-factorizable
radiative corrections we know that they disappear 
if the integration over invariant masses is performed [4-6]. As the 
integration over invariant masses restores the stable particle scenario, 
the absence of the contribution due to non-factorizable corrections
in the integrated quantities is
in accordance with the non-observability of the  
Coulomb phase in the familiar theories with the stable 
particles.   

As for the cancellation of the virtual photon poles against 
the real emission 
we argue that this cancellation occurs only if the integration
over invariant mass of  
at least one of the resonances is 
performed. 

Let us give a summary of the formulae presented in the text:
\begin{itemize}
\item{Non-factorizable radiative correction to the differential 
cross-section for a single resonance production is given by 
eq.(\ref{single}).}
\item{Non-factorizable corrections for the matrix elements describing  
production of two
resonances are given by:
\begin{enumerate}
\item{ three-point function -- eq.(\ref{tria});}
\item{ four-point function -- eq.(43);}
\item{ five-point function -- eq.(\ref{five}).}
\end{enumerate}
}\end{itemize}

From the phenomenological side 
our study is motivated by a future 
investigation   
of heavy unstable particles. The vivid example is 
provided by the study of the reaction $e^-e^+ \to W^-W^+$ 
at LEP 2. It seems that the planning 
accuracy of the measurement
and the proposed technique of measuring the line shape of the invariant 
mass distribution requires taking into account QED non-factorizable 
corrections as well. 

As the dominant contribution to $e^-e^+ \rightarrow W^-W^+$ comes 
from the $t$-channel neutrino exchange, on the first glance it seems 
that the six-point function is actually needed for this case. However,
simple estimates show that practically for the whole phase-space of the 
final particles  
factorization of the Born amplitude is still valid. Hence it is 
sufficient to use the five-point function for the 
description of the production of two $W$ bosons at LEP2.

We note that the energy region for the LEP 2 ($\sqrt{s} = 170 - 200$ GeV)
 is the intermediate but not really threshold energy region. 
Consequently  one has to consider the effects of the 
final state interaction between 
decay products of the resonances as well: it is likely that 
the Coulomb correction alone (which 
is definitely the leading one in the threshold region) is not 
sufficient for the LEP2 energy region.

As it has been indicated above our results beeing obtained in the soft
approximation are universal. For the illustrative purposes we apply them to 
the process $e^+e^- \to W^+W^- \to e^+ \nu _e e^- \bar \nu _e$.
The values of different sources of virtual contributions as well as 
corresponding pieces in real interference are presented in the figs.5--6. 

We stress however that these {\it numerical}
consequences of our results for the observable quantities  seem 
to  depend strongly on the experimental procedure which will be 
used in the real life experiments. It must be clear from the above 
discussion that the cancellation of 
the soft real emission against the virtual corrections is quite delicate 
in the case of the production of the unstable particles. 
Therefore a more realistic treatment of the 
soft ($\omega \sim \Gamma$)radiation is necessary. 
This is basically the main reason why 
we do not see much sense in an exhaustive numerical analyses of our 
formulae. 

Another phenomenological issue which we mention here is the 
possibility to measure the invariant mass distribution of the 
top quark at $e^+e^-$ and $\gamma \gamma$ colliders. Our 
formulae can be also applied
for the $O(\alpha _s)$ non-factorizable corrections in this case  
in the spirit of \cite{Mel}, \cite{Sum}. 
Let us note however that in this case the influence of the hadronization on 
the precise determination of the top mass should be considered.
The discussion of this important issue can be found in the 
ref. \cite {Khoze}.

To conclude, we want to emphasize once more that the non-factorizable 
corrections change the shape of the invariant-mass 
distribution while preserve the total probability [4-6]. 
  As the study of the properties  of the
unstable fundamental particles requires the 
measurement of the invariant mass distributions our results must be taken 
into account while preparing for the analysis of the forthcoming 
high-precision experiments on the unstable particle production. Not only 
high statistics will be important but also our possibilities to 
make the correct correspondence between 
the results of the perturbative
calculations of the masses and widths of the unstable particles in the
framework of the  
Standard 
Model with the experimentally measured quantities.       

\section {Acknowledgments}
We are indebted to Prof. J.~G.~K\"{o}rner for  
support and encouragement in the course of this work.
We would like to thank Prof. I.~F.~Ginzburg for 
a number of useful discussions. We are grateful 
to referees of journal of {\em Nucl. Phys. B },
whose criticism has been decisive in making this manuscript intelligible.
K.M. is grateful to the Graduiertenkolleg ``Elementarteilchenphysik''
of the Mainz University for the support. 
O.I.Y. acknowledges support of the Deutsche Forschung Gemeinschaft.

\vspace{0.5cm}
{\Large \bf Appendix}

\vspace{0.5cm}                                
Let us introduce the following integral:
\be
F_1(a,\beta|x_i)= \int \limits _{-1}^{1} \frac {dx}{x-a} \log(1+\beta x)
(\theta (x-x_i)- \theta (x_i-x)).  
\ee
Here $a$ is a complex number with the non-zero imaginary part.  
Also $x_i$ is an arbitrary number 
satisfying $ -1 \le x_i \le 1 $. The result of the integration then 
reads:  
\begin{eqnarray}
F_1(a,\beta|x_i)=-2\log\Big (\frac {(a-x_i)\beta}{1+\beta a}\Big )
\log(1+\beta x_i)
&+&\log\Big (\frac {(1+a)\beta }{1+\beta a}\Big )\log(1-\beta)+ \nonumber \\
\log\Big (\frac {(a-1)\beta }{1+\beta a}\Big )\log(1+\beta) 
-2\Li _2\Big (\frac {1+\beta x_i}{1+a\beta}\Big )
&+&\Li _2\Big (\frac {1-\beta}{1+a\beta}\Big )        
+\Li _2 \Big (\frac {1+\beta}{1+a\beta}\Big ). \nonumber
\end{eqnarray}
In our formulas we also need this function in the case when $a$
is real but in some restricted cases, namely $a=x_i$. In this special case 
the divergence is of the collinear origin and we regularize it keeping the
mass of the light particle finite. 
Hence the result for this function with $a=x_i$ reads:
\begin{eqnarray}
F_1(x_i,\beta|x_i)=\log(1+\beta x_i)
\log\Big (\frac {4{E_i}^2}{{m_i}^2}\Big )
-\Li _2\Big (\frac {(1+x_i)\beta}{1+x_i\beta} \Big )        
-\Li _2 \Big (\frac {(x_i-1)\beta}{1+x_i\beta} \Big ). \nonumber
\end{eqnarray}
\par
Our next function is defined as following:
\be
F_2(a|x_i)= \int \limits _{-1}^{1} \frac {dx}{x-a}
(\theta (x-x_i)- \theta (x_i-x)).  
\ee  
The result of the integration is:
\be
F_2(a|x_i)=-2\log(x_i-a)+\log(-1-a)+\log(1-a).
\ee
When $a=x_i$ this function equals:
\be
F_2(x_i|x_i)=\log\Big (\frac {4{E_i}^2}{{m_i}^2}\Big ).
\ee

Next we present the quantities necessary for the eqs.(57-59):
\be
R_{4}=\frac {2\beta x_4 }{m_t^2\xi (x_4)},~~~
R_{i=+,-}=\frac {-\beta x_i}{m_t^2\xi (x_i)},
\ee
$$
R_{\xi,4}=\frac {2\beta (D_1+D_2)(K_4(D_1+D_2)+\beta (D_1-D_2)N_4)}
{m_t ^6(1+x_{34})
\xi (x_4) \xi (x_+) \xi (x_-)}
$$
where all notations are the same as in the main text of the paper.
The quantities for the index $3$ can be obtained from 
the previous ones by 
direct substitution  $4 \rightarrow 3 $.

\newpage

\begin{figure}[htb]
\epsfxsize=12cm
\centerline{\epsffile{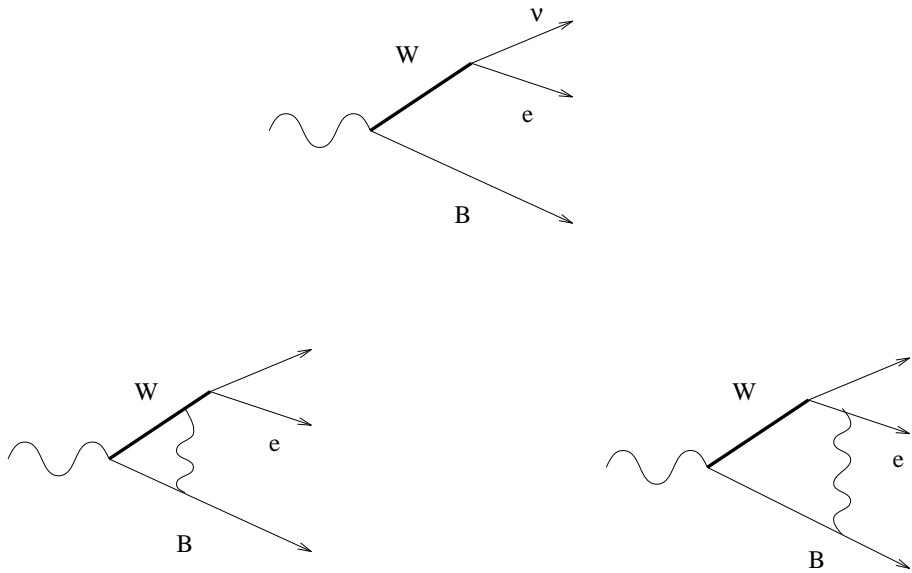}}
\caption[]{Born graph  and
graphs responsible for the non-factorizable corrections for the simple 
model (see sect.2).}
\end{figure}

\begin{figure}[htb]
\epsfxsize=12cm
\centerline{\epsffile{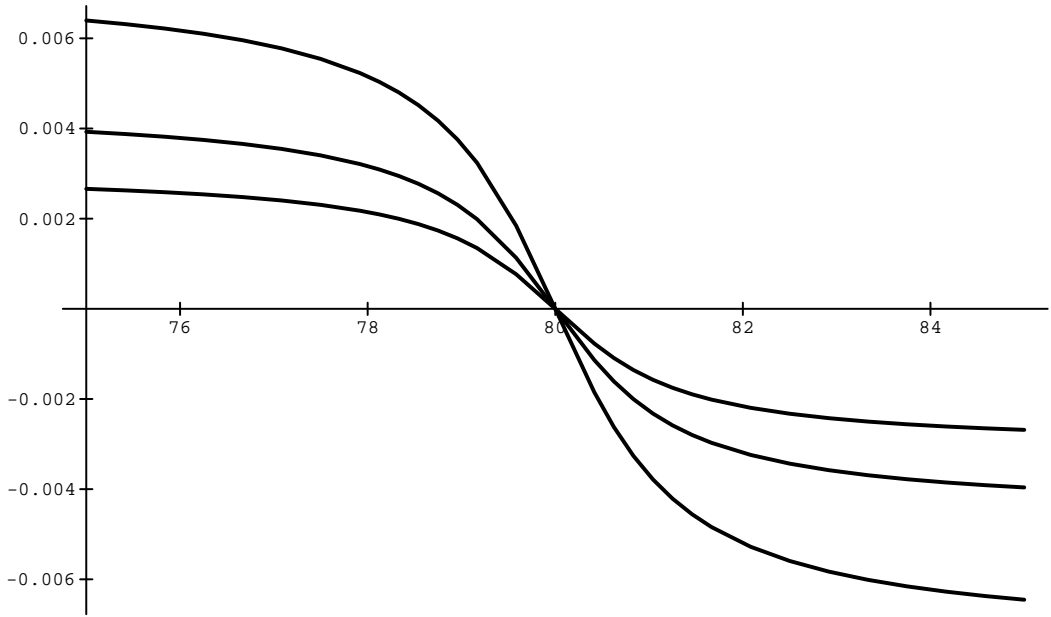}}
\caption[]{The relative size of the non-factorizable
radiative corrections in the simple model (see eq.(11) with $\eta$ from 
eq.(15)). Curves $A$, $B$,
$C$ correspond to the total energies $\sqrt {s} = 180,~190,~200$ GeV
respectively. We use $m_W =80$ GeV and $\alpha = 1/137$.}
\end{figure}

\begin{figure}[htb]
\epsfxsize=15cm
\centerline{\epsffile{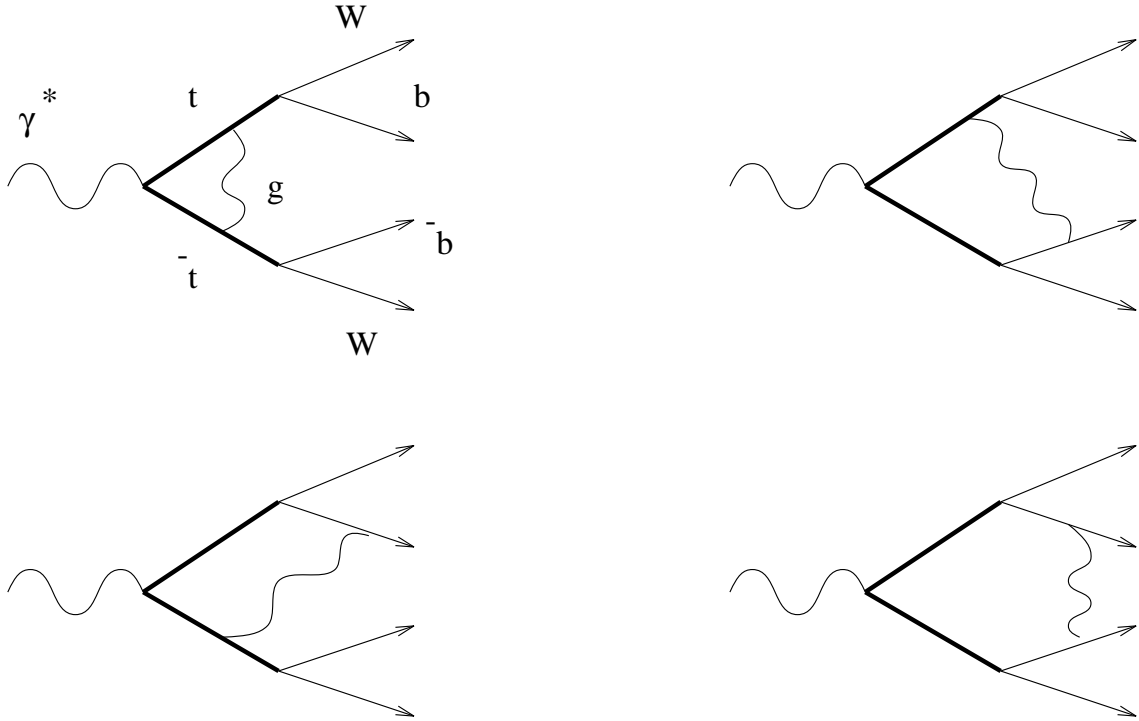}}
\caption[]{Non-factorizable graphs for the process $\gamma ^* \to t\bar t
 \to bW^+\bar b W^-$.}
\end{figure}

\begin{figure}[htb]
\epsfxsize=15cm
\centerline{\epsffile{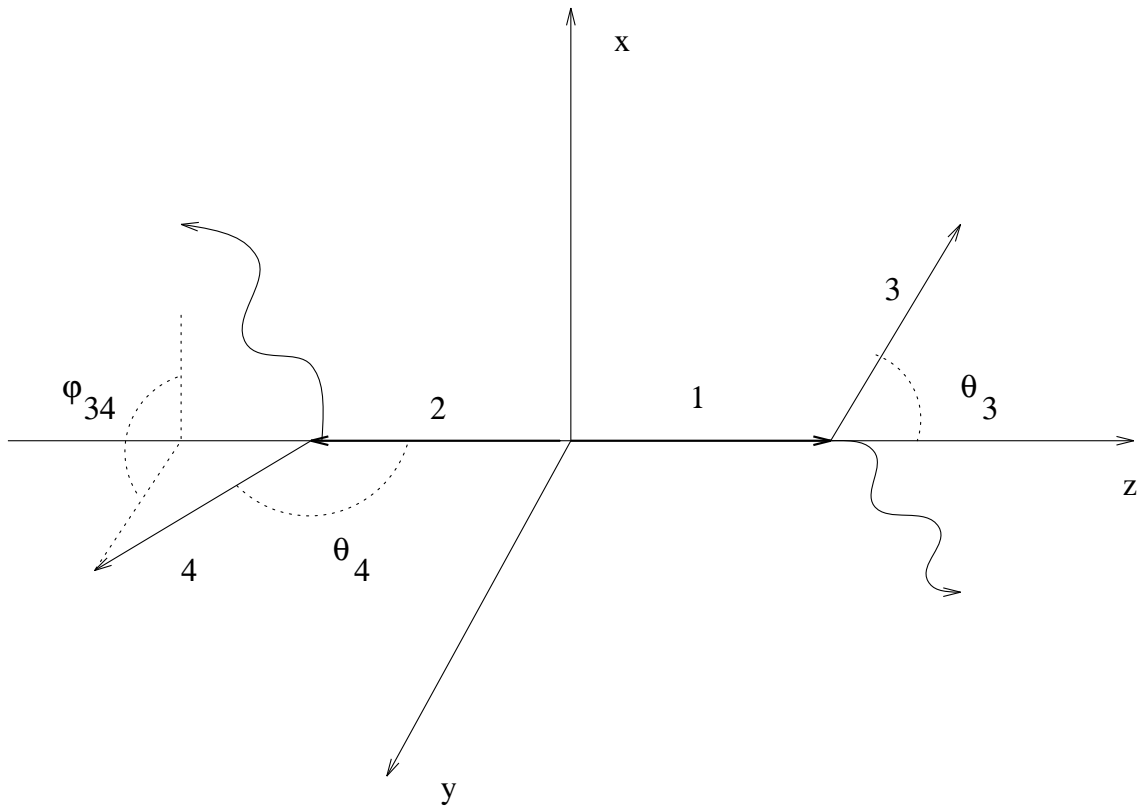}}
\caption[]{Geometry of the discussed reactions.}
\end{figure}

\begin{figure}[htb]
\epsfxsize=12cm
\centerline{\epsffile{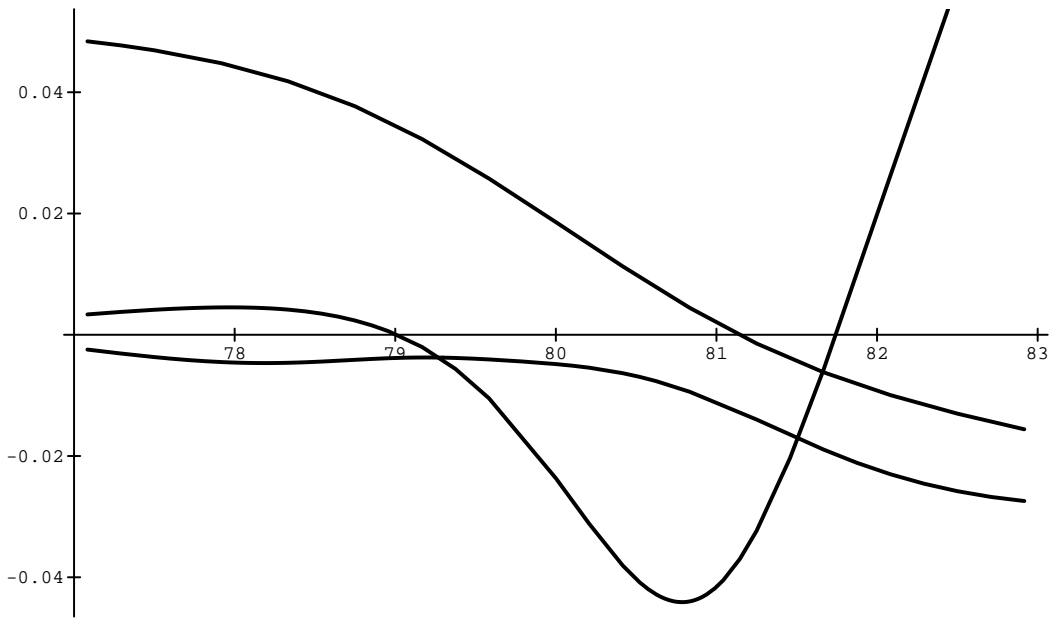}}
\caption[]{Relative non-factorizable corrections to completely 
differential cross sectionon of the process 
$e^+e^-\to W^+W^-\to e^+\nu_e e^-\bar\nu_e$ 
as a function of invariant mass of the $e^+ \nu _e$ system $m_2$ in GeV 
for the fixed invariant mass of $e^- \bar \nu _e$ $m_1=78$ GeV.
We use $\sqrt{s} = 180$ GeV, $m_W = 80$ GeV, 
 $\alpha =1/137$, $\theta _{W-e^-} = 30 ^{\circ},~ 
\theta _{W^-e^+} = 150 ^{\circ},~~ \varphi _{e^+e^-} =0$. 
Curves A, B, C correspond to the conrtibutions due to three--,
 four--, and five--point functions respectively.}
\end{figure}

\begin{figure}[htb]
\epsfxsize=12cm
\centerline{\epsffile{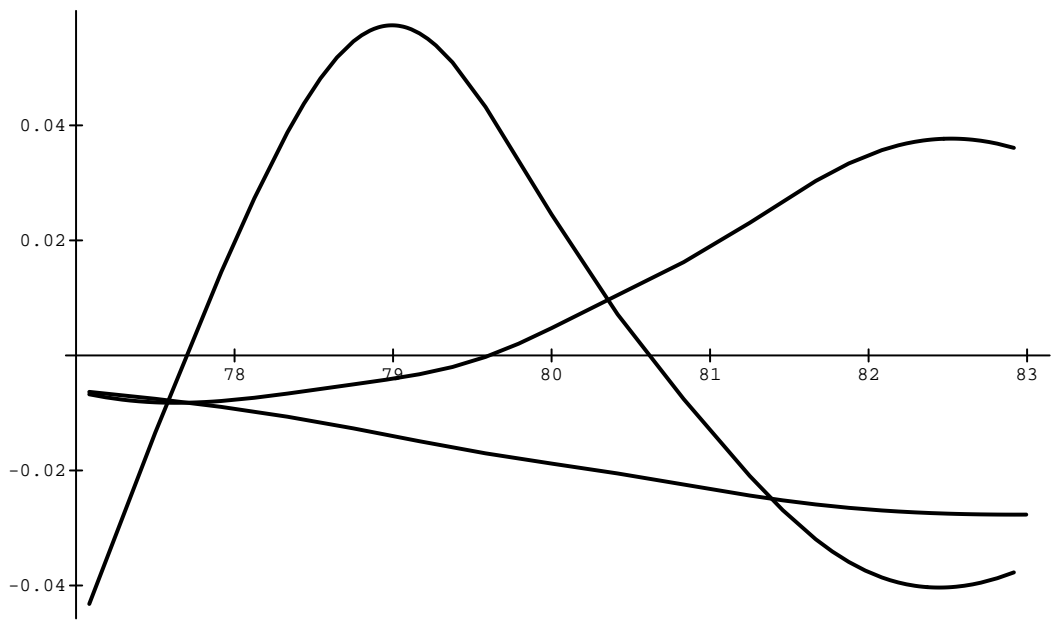}}
\caption[]{The same as in fig.4, but for $m_1=82$ GeV.}
\end{figure}
\end{document}